\documentclass[11pt,a4paper,english,nofootinbib]{revtex4}
\usepackage{lmodern}

\usepackage[T1]{fontenc}
\usepackage[latin9]{inputenc}
\usepackage{babel}

\usepackage{amsmath}
\usepackage{graphicx}
\usepackage{amssymb}
\usepackage{esint}
\usepackage[unicode=true, pdfusetitle,
 bookmarks=true,bookmarksnumbered=false,bookmarksopen=false,
 breaklinks=false,pdfborder={0 0 1},backref=false,colorlinks=false]
 {hyperref}

\makeatletter
\@ifundefined{textcolor}{}
{%
 \definecolor{BLACK}{gray}{0}
 \definecolor{WHITE}{gray}{1}
 \definecolor{RED}{rgb}{1,0,0}
 \definecolor{GREEN}{rgb}{0,1,0}
 \definecolor{BLUE}{rgb}{0,0,1}
 \definecolor{CYAN}{cmyk}{1,0,0,0}
 \definecolor{MAGENTA}{cmyk}{0,1,0,0}
 \definecolor{YELLOW}{cmyk}{0,0,1,0}
 }

\usepackage{latexsym}\usepackage{bm}

\makeatother

\makeatother

\begin{document}

\title{Higgs Mechanism  for  New Massive Gravity and Weyl Invariant Extensions of Higher Derivative Theories  }

\author{Suat Dengiz}

\email{e169656@metu.edu.tr}

\affiliation{Department of Physics,\\
 Middle East Technical University, 06531, Ankara, Turkey}

\author{Bayram Tekin}

\email{btekin@metu.edu.tr}

\affiliation{Department of Physics,\\
 Middle East Technical University, 06531, Ankara, Turkey}

\date{\today}
\begin{abstract}
New Massive Gravity  provides a  non-linear extension of the Fierz-Pauli mass for gravitons in $2+1$ dimensions. Here  we construct a  Weyl invariant version of this theory.
When the Weyl symmetry is broken, the graviton gets a mass  in analogy with the  Higgs mechanism. In (anti)-de Sitter backgrounds, the symmetry can be broken spontaneously, but in flat backgrounds radiative corrections, at the two loop level,  break the Weyl symmetry \`a la Coleman-Weinberg mechanism.  We also construct the Weyl invariant extensions of some other higher derivative models, such as the Gauss-Bonnet theory ( which reduces to the Maxwell theory in three dimensions ) and the Born-Infeld type gravities.

 \tableofcontents{}\[
\]

\end{abstract}
\maketitle
\section{Introduction}

New Massive gravity [NMG] \cite{BHT1,BHT2} defined by the action
\begin{equation}
 I_{NMG}=\frac{1}{\kappa^2} \int d^3 x \sqrt{-g} \Big [\sigma R-2 \lambda  m^2 + \frac{1}{m^2} \Big ( R^2_{\mu\nu}-\frac{3}{8}R^2 \Big ) \Big ],
\label{nmg}
\end{equation}
describes, at the linearized level, a massive graviton with 2 degrees of freedom both around its flat (for $\lambda = 0$)  and its (anti)-de Sitter  ( $\lambda > -1$ )
vacua.  The theory is unitary at the tree level for certain choices and ranges of parameters, the details of which, and some related work on the perturbative spectrum of the model, can be found in \cite{BHT2,GulluTekin,deser,nakasone,liusun,canonical,cubic}. There are at least two reasons why NMG is a valuable theoretical lab for ideas about "quantum gravity" : First of all,  it is a super-renormalizable theory \cite{stelle} and secondly, which is more relevant to the current work, it provides a non-linear extension to the (unitary)  Fierz-Pauli mass of a graviton.  While all this is quite interesting, there is a missing part of the mass-puzzle here: The graviton, after linearization, acquires a hard mass, instead, one would like to understand the mass  arising from a symmetry breaking  Higgs-type  mechanism, just like one understands the mass of the various fields, such as the non-abelian vector bosons {\it etc}. 
In this paper, we provide such a mechanism by finding the Weyl invariant extension of NMG and by showing that the vacuum of the theory is not Weyl invariant. How the symmetry breaking takes  place, assuming one does not break it by hand, crucially depends on whether one is working around (A)dS or flat backgrounds. As we shall see, the latter is somewhat more intricate.  Before we discuss the Weyl invariant extension of NMG by introducing a Weyl gauge field and a scalar field, let us note that,   it was  realized in  \cite{deser} that the linearized form of the quadratic part of  NMG  is Weyl invariant and the linearized part of the Einstein-Hilbert term breaks this symmetry.   Taking this observation as a hint, we will make the {\it  full action }, not just the linearized one, Weyl invariant and find the field equations and show that the scalar field develops a vacuum expectation value in the case of (A)dS backgrounds. For flat backgrounds, Weyl symmetry is broken at the two loop level. 

Besides the Weyl invariant extension of NMG, we will  give Weyl invariant extensions of the Einstein-Gauss-Bonnet (EGB) theory in generic dimensions. Specifically, for $2+1$ dimensions, the Weyl-invariant GB combination  reduces just to a Maxwell  term (with a compensating scalar field), which is interesting since the non-Weyl invariant GB combination identically vanishes in this dimension. We also give the  Weyl invariant versions of the Born-Infeld-NMG  action \cite{ BINMG}  which was constructed to reproduce NMG at the quadratic expansion in curvature and the theory found in \cite{Sinha} using the existence of holographic c-functions , at the cubic and quartic expansions in curvature .

The lay out of the paper is as follows: In Section II, we start with a brief review of constructing Weyl invariant actions and the relevant Weyl invariant tensors. In Section III,  we give a Weyl invariant extension of NMG and the field equations and study the maximally symmetric vacua.  In section IV,  we discuss the Weyl invariant form of EGB theory and the BINMG theory. 

 \section{Basics of Weyl invariance}

Weyl's original idea ( see \cite{oraif} for a brief review) was to unify gravity and electromagnetism by "gauging" the metric as $ g_{\mu \nu} \rightarrow  e^{ \int A_\alpha dx^\alpha} g_{\mu \nu}$, where $A_{\mu}$ is the vector potential. While this procedure did not give a correct unified theory, the idea of  coupling  electromagnetism to charged fields through the "gauging" of the wave functions  survived. In a Poincare invariant theory, Weyl's idea boils down to upgrading the rigid scale invariance\footnote {That is setting  $x^\mu \rightarrow  \lambda x^\mu$   and similarly scaling the fields  with their scale dimensions $d$ as $\varphi \rightarrow \lambda^d \varphi$, where $\lambda$ 
is a constant.}   to a local  scale invariance. The method of how the Weyl-gauging idea is  implemented in a given theory is important, since, the theories we shall consider are higher derivative rather complicated models, therefore an economical way of writing their Weyl invariant version   is required. We start with the  basics.  We will work in a generic  $n$ dimensional space-time and specify to $n=2+1$ later.  Perhaps the best way to introduce the idea is to start with the kinetic part of the scalar field action 
\begin{equation}
S_{\Phi}=- \frac{1}{2}\int d^n x \sqrt{-g} \partial_\mu \Phi  \partial_\nu \Phi g^{\mu \nu}.
\end{equation}
To make this action Weyl invariant in the background of dynamical gravity, one should have invariance under
\begin{equation}
 g_{\mu\nu} \rightarrow g^{'}_{\mu\nu}=e^{2 \lambda(x)} g_{\mu\nu}, \hskip 2 cm \Phi \rightarrow \Phi^{'} =e^{-\frac{(n-2)}{2}\lambda(x)} \Phi,
\end{equation}
where $ \lambda(x) $ is an arbitrary function of the coordinates and the derivatives should be replaced with the (real) gauge covariant derivatives (not to be confused with the spacetime covariant derivative $\nabla_\mu$ that we shall use below)  
\begin{equation}
 \mathcal{D}_\mu \Phi =\partial_\mu \Phi -\frac{n-2}{2} A_\mu \Phi  , \hskip 2 cm  \mathcal{D}_\mu g_{\alpha \beta} =\partial_\mu g_{\alpha\beta} + 2 A_\mu g_{\alpha \beta},
\end{equation}
where $ A_\mu $  the  Weyl's  gauge field which transforms as
\begin{equation}
 A_\mu \rightarrow A^{'}_\mu = A_\mu - \partial_\mu \lambda(x).
\end{equation}
By construction, one then has 
\begin{equation}
 ( \mathcal{D}_\mu g_{\alpha \beta})^{'}=e^{2 \lambda(x)} \mathcal{D}_\mu g_{\alpha \beta},\hskip 2 cm   (\mathcal{D}_\mu \Phi)^{'}=e^{-\frac{(n-2)}{2}\lambda(x)} \mathcal{D}_\mu \Phi. 
\end{equation}
The field strength $F_{\mu \nu} = \partial_\mu A_\nu - \partial_\nu A_\mu $ is gauge invariant, but the Maxwell-type action needs a compensating Weyl scalar
\begin{equation}
S_{A^\mu} =  - \frac{1}{2} \int d^n x \sqrt{-g}\,\, \Phi^{\frac{2(n-4)}{n-2}} F_{\mu \nu} F^{\mu \nu}.
\label{maxwell}
\end{equation}
As for the gravity side, the shortest way to implement Weyl invariance is to define a Weyl invariant connection using the Christoffel connection and compute the Riemann and Ricci tensors and the Ricci scalar from that. Quite easily one can see that the following does the job
\begin{equation}
 \tilde{\Gamma}^\lambda_{\mu\nu}=\frac{1}{2}g^{\lambda\sigma} \Big ( \mathcal{D}_\mu g_{\sigma\nu}+\mathcal{D}_\nu g_{\mu\sigma}
-\mathcal{D}_\sigma g_{\mu\nu} \Big ).
\end{equation}
Then the Weyl-invariant " Riemann tensor" can be computed as\footnote{One could allow  the gauge covariant derivative act on the gauge field as  $\mathcal{D}_{\mu} A_\nu \equiv \partial_\mu A_\nu - A_\mu A_\nu$ and define the combination of gauge and metric covariant derivatives as ${\tilde{\mathcal{D}}}_{\mu} A_\nu \equiv \nabla_\mu A_\nu - A_\mu A_\nu$ , which somewhat  simples the subsequent computations. }
 \begin{equation}
\begin{aligned}
  \tilde{R}^\mu{_{\nu\rho\sigma}} [g,A]&=\partial_\rho \tilde{\Gamma}^\mu_{\nu\sigma}-\partial_\sigma \tilde{\Gamma}^\mu_{\nu\rho}
+ \tilde{\Gamma}^\mu_{\lambda\rho} \tilde{\Gamma}^\lambda_{\nu\sigma}-\tilde{\Gamma}^\mu_{\lambda\sigma} \tilde{\Gamma}^\lambda_{\nu\rho} \\
& =R^\mu{_{\nu\rho\sigma}}+\delta^\mu{_\nu}F_{\rho\sigma}+2 \delta^\mu{_[\sigma} \nabla_{\rho]} A_\nu 
+2 g_{\nu[\rho}\nabla_{\sigma]} A^\mu \\
& \quad +2 A_[\sigma   \delta_{\rho]}\,^\mu A_\nu  
+2 g_{\nu[\sigma}  A_{\rho]} A^\mu  +2 g_{\nu[\rho} \delta_{\sigma]}\,^\mu  A^2 , 
\end{aligned} 
\end{equation}
where we have used the notation $ 2 A_{[ \rho} B_{\sigma]} \equiv A_\rho B_\sigma -  A_\sigma B_\rho$ and $A^2= A_\mu A^\mu$. Note that we do not insist on the original symmetries of the Riemann tensor, we accept what we get from the above construction, since, at the end, we would like to  get a Weyl invariant action and once we do that, we will get back to the original  fields.  Similarly, the Weyl-invariant Ricci tensor reads  
\begin{equation}
\begin{aligned}
\tilde{R}_{\nu\sigma} [g,A]&= \tilde{R}^\mu{_{\nu\mu\sigma}}[g,A] \\
&=R_{\nu\sigma}+F_{\nu\sigma}-(n-2)\Big [\nabla_\sigma A_\nu - A_\nu A_\sigma +A^2  g_{\nu\sigma} \Big ]-  g_{\nu\sigma}\nabla \cdot A,
\end{aligned}
\end{equation}
where $\nabla \cdot A \equiv \nabla_\mu  A^\mu$. One more contraction gives the scalar curvature,
\begin{equation}
 \tilde{R}[g,A]=R-2(n-1)\nabla \cdot A-(n-1)(n-2) A^2,
\end{equation}
 which is {\it not} Weyl invariant, but transforms as $ (\tilde{R}[g,A])^{'} = e^{-2 \lambda (x) }  \tilde{R}[g,A]$. Therefore, to write the Weyl-invariant version of the Einstein-Hilbert action, we need a compensating Weyl scalar 
 \begin{equation}
S= \int d^n x \sqrt{-g} \Phi^2  \tilde{R}[g,A] = \int d^n x \sqrt{-g} \Phi^2\Big (R-2(n-1)\nabla \cdot A-(n-1)(n-2) A^2 \Big).
\label{eh}
\end{equation}
What is interesting about this action is that, suppose, one does not add dynamics to the gauge field, then, independent of whether one adds an explicit kinetic term to the scalar field or not,  after eliminating the gauge field by using its field equation
\begin{equation}
A_{\mu }= \frac{2}{n-2}\partial_{\mu} \ln \Phi,
\end{equation} 
one ends up with the conformally coupled scalar-tensor theory
\begin{equation}
S=\int d^n x \sqrt{-g}\Big ( \Phi^2 R+4\frac{(n-1)}{n-2} \partial_\mu \Phi \partial^\mu \Phi  \Big).
\end{equation}
Of course, in higher curvature theories, that we shall discuss, generically, the gauge field will necessarily be dynamical.  If the scalar field takes the constant value (inverse of the Newton's constant) then General Relativity without a cosmological constant is recovered.  To introduce a cosmological constant (in this limit)  one should add a Weyl-invariant potential to the scalar field yielding 
\begin{equation}
S_\Phi=- \frac{1}{2}\int d^n x \sqrt{-g}\Big (   \mathcal{D}_\mu \Phi \mathcal{D}^\mu\Phi +\nu \, \Phi^{\frac{2n}{n-2}}\Big ) ,
\label{scalarwithpot}
\end{equation}
where $\nu \ge 0$ is a dimensionless coupling constant. As expected, at least in flat backgrounds, this yields a renormalizable theory in $n=3$ and $n=4$. (Here we do not discuss the rather special case of $n=2$ ). After this excursion to the Weyl gauging in Einstein's gravity, let us study the higher derivative models. 

\section{Weyl-invariant New Massive Gravity}

Using the tools developed in the previous section, the Weyl-invariant extension of  NMG can be written as 
\begin{equation}
  \tilde{S}_{NMG}= \int d^3x \sqrt{-g} \bigg [\sigma \Phi^2 \tilde{R}
+ \Phi^{-2} \Big ( \tilde{R}^2_{\mu\nu}-\frac{3}{8}\tilde{R}^2 \Big ) \bigg ]  + S_\Phi,
\label{WNMG}
\end{equation}
where  the last term is the scalar action given in (\ref{scalarwithpot}). We could also add the "Maxwell'' term  (\ref{maxwell}) but, in any case, it will be generated as one can see from the explicit form of the action 
\begin{equation}
\begin{aligned}
  \tilde{S}_{NMG}= \int d^3 x \sqrt{-g} & \bigg \{\sigma \Phi^2 \Big(R-4\nabla \cdot A -2A^2  \Big) \\
&+\Phi^{-2} \bigg [R^2_{\mu\nu}-\frac{3}{8} R^2-2 R^{\mu\nu}\nabla_\mu A_\nu+ 2R^{\mu\nu}A_\mu A_\nu \\
& \qquad \quad  +R\, \nabla \cdot A-\frac{1}{2}R A^2+2  F_{\mu\nu}^2 +(\nabla_\mu A_\nu)^2 \\
& \qquad \quad-2 A_\mu A_\nu \nabla^\mu A^\nu-(\nabla \cdot A)^2+\frac{1}{2}A^4 \bigg ] \bigg \}  +S_\Phi.
\label{winmg}
\end{aligned}
\end{equation}
From a formal point of view, if we just set the gauge field to zero and  choose $\Phi= \sqrt{m}$ and   $\nu = 2 \lambda$, then we get the NMG action (\ref{nmg}) with a fixed coupling constant $ \kappa = m^{-1/2}$.  With just one scalar field, one can of course not generate two different scales, therefore in the Weyl-invariant extension of NMG, the three dimensional Newton's constant is related to the mass of the graviton when the scalar field freezes.  It would be much more interesting  if one can show that this formal limit of Weyl-invariant NMG arises as the vacuum solution of the theory.  This means that the vacuum should break the Weyl symmetry. [We can add an explicit Weyl symmetry breaking term, such as the mass of the scalar field as was done in \cite{deserscale} for the case of the conformally coupled scalar field in four dimensions, but we will not do that. We would like to keep the analogy with the Higgs mechanism and show that the Weyl symmetry breaks spontaneously
   (not explicitly) for AdS backgrounds and  at the two-loop level for flat backgrounds.]  Therefore, we need the field equations. Not to clutter the notation let us keep the part of the equations coming from the scalar field in the compact form as  $\frac{ \delta S_{\Phi}}{\delta \Phi}$ {\it etc}.  
Then, from the variation of (\ref{winmg}) with respect to $ \delta g^{\mu \nu}$, after a tedious computation,  disregarding the boundary terms, one obtains
\begin{equation}
 \begin{aligned}
 & \sigma \Phi^2 G_{\mu\nu}+\sigma g_{\mu\nu} \Box \Phi^2-\sigma \nabla_\mu \nabla_\nu \Phi^2-4 \sigma \Phi^2 \nabla_\mu A_\nu+2 \sigma  g_{\mu\nu}\Phi^2 \nabla \cdot A
-2 \sigma \Phi^2 A_\mu A_\nu \\
& + \sigma g_{\mu\nu} \Phi^2 A^2 +2 \Phi^{-2} [R_{\mu\sigma\nu\alpha}-\frac{1}{4} g_{\mu\nu}R_{\sigma\alpha}]R^{\sigma\alpha}
+\Box (\Phi^{-2} G_{\mu\nu})+\frac{1}{4}[g_{\mu\nu}\Box - \nabla_\mu \nabla_\nu]\Phi^{-2}R  \\
&+g_{\mu\nu}G^{\sigma\alpha}\nabla_\sigma \nabla_\alpha \Phi^{-2} 
-2 G^\sigma{_\nu}\nabla_\sigma \nabla_\mu \Phi^{-2}-2(\nabla_\mu G^\sigma{_\nu})(\nabla_\sigma \Phi^{-2} )+\frac{3}{16}g_{\mu\nu}\Phi^{-2} R^2 \\
&-\frac{3}{4}\Phi^{-2}R R_{\mu\nu} +g_{\mu\nu}\Phi^{-2}R_{\alpha\beta} \nabla^\alpha A^\beta 
-2\Phi^{-2}R_{\alpha\nu} \nabla_\mu A^\alpha -2\Phi^{-2}R_{\beta\mu} \nabla^\beta A_\nu-\Box(\Phi^{-2}\nabla_\mu A_\nu) \\
& -g_{\mu\nu}\nabla_\beta \nabla_\alpha(\Phi^{-2}\nabla^\alpha A^\beta ) 
+\nabla_\alpha \nabla_\nu (\Phi^{-2}\nabla^\alpha A_\mu ) +\nabla_\beta \nabla_\nu (\Phi^{-2}\nabla_\mu A^\beta) 
-g_{\mu\nu}\Phi^{-2}R^{\alpha\beta} A_\alpha A_\beta \\
&+4\Phi^{-2}R_{\alpha\nu} A_\mu A^\alpha+ \Box(\Phi^{-2}A_\mu A_\nu) 
-2\nabla^\alpha \nabla_\nu(\Phi^{-2} A_\alpha A_\mu)+g_{\mu\nu}\nabla^\alpha \nabla^\beta(\Phi^{-2} A_\alpha A_\beta) \\
& +\Phi^{-2}G_{\mu\nu}\nabla \cdot A+g_{\mu\nu} \Box(\Phi^{-2} \nabla \cdot A  )-\nabla_\mu \nabla_\nu (\Phi^{-2}  \nabla \cdot A )
+ \Phi^{-2} R \, \nabla_\mu A_\nu-\frac{1}{2} \Phi^{-2}G_{\mu\nu}A^2 \\
& -\frac{1}{2} g_{\mu\nu}\Box (\Phi^{-2}A^2)+ \frac{1}{2} \nabla_\mu \nabla_\nu(\Phi^{-2}A^2)-\frac{1}{2}\Phi^{-2} R A_\mu A_\nu
-\Phi^{-2}[g_{\mu\nu}F_{\alpha\beta}^2+4 F_\nu{^\alpha}F_{\alpha\mu}] \\
&-\frac{1}{2}g_{\mu\nu}\Phi^{-2}(\nabla_\alpha A_\beta)^2+\Phi^{-2}\nabla_\mu A_\alpha \nabla_\nu A^\alpha+\Phi^{-2}\nabla_\beta A_\nu \nabla^\beta A_\mu 
 +g_{\mu\nu}\Phi^{-2} A^\alpha A^\beta \nabla_\alpha A_\beta \\
&-2 \Phi^{-2} A_\nu A^\alpha \nabla_\mu A_\alpha-2 \Phi^{-2} A_\mu A^\beta \nabla_\beta A_\nu+\frac{1}{2}g_{\mu\nu}(\nabla \cdot A )^2 
-2 (\nabla \cdot A) \nabla_\mu A_\nu \\
& -\frac{1}{4} g_{\mu\nu}\Phi^{-2} A^4 +\Phi^{-2} A_\mu A_\nu A^2= - \frac{1}{\sqrt{-g}}\frac {\delta S_{\Phi}}{ \delta g^{\mu \nu}}.
\end{aligned}
\end{equation}
Variation with respect to $ \delta \Phi$ yields 
\begin{equation}
\begin{aligned}
 &2\sigma \Phi \Big(R-4\nabla \cdot A-2A^2 \Big)-   2\Phi^{-3}\bigg [R^2_{\mu\nu}-\frac{3}{8} R^2
-2 R^{\mu\nu}\nabla_\mu A_\nu+ 2R^{\mu\nu}A_\mu A_\nu 
 +R\, \nabla \cdot A -\frac{1}{2}R A^2 \\
&+2 F_{\mu\nu}^2+ (\nabla_\mu A_\nu)^2-2 A_\mu A_\nu \nabla^\mu A^\nu-(\nabla \cdot A)^2+\frac{1}{2}A^4 \bigg ]= - \frac{1}{\sqrt{-g}}\frac{ \delta S_{\Phi}}{\delta \Phi}.
\label{phidenklemi}
\end{aligned}
\end{equation}
Finally, the Weyl gauge field equation reads
\begin{equation}
\begin{aligned}
& -4 \nabla_\mu \Phi^2 + 4 \Phi^2 A_\mu+2 R^\nu{_\mu}\nabla_\nu \Phi^{-2}+4 R_{\mu\nu}A^\nu
-R \nabla_\mu \Phi^{-2}-\Phi^{-2}RA_\mu+8 \nabla^\nu(\Phi^{-2} \nabla_\mu A_\nu ) \\
&-10 \nabla^\nu(\Phi^{-2} \nabla_\nu A_\mu )+2 \nabla_\alpha (\Phi^{-2}A^\alpha A_\mu)
-2 \Phi^{-2} (\nabla_\mu A_\nu)A^\nu -2 \Phi^{-2} (\nabla_\nu A_\mu)A^\nu \\ 
&+2 \nabla_\mu (\Phi^{-2}\nabla \cdot A )+2 A_\mu A^2= -\frac{1}{\sqrt{-g}} \frac{ \delta S_{\Phi}}{\delta A^{\mu}}.
\label{amudenklemi}
\end{aligned}
\end{equation}
Let us now consider the vacuum solution to these equations for the case when the spacetime is (anti)-de Sitter. Not to break the local  Lorentz invariance of the vacuum,  let us set 
$F_{\mu \nu}=0$ and choose the gauge for which $A_\mu =0$. This is in fact the most sensible ansatz to take for the Weyl gauge field part. Then let 
\begin{equation}
\Phi \equiv \sqrt{m},  \hskip  2 cm  R_{\mu \nu} = 2 \Lambda  g_{\mu \nu} , 
\end{equation}  
 In three dimensions once Ricci tensor is given, Riemann tensor is fixed, so we do not depict it separately.  The field equations will relate  $m$ and $\Lambda$.  The gauge field equation (\ref{amudenklemi}) is automatically satisfied.  The other two equations give the same equation
\begin{equation}
\nu \, m^4 - 4 \sigma m^2 \Lambda-\Lambda^2  =0.
\label{vac}
\end{equation}
 Let us first consider the $\nu > 0$ case. The discussion bifurcates: One can  either assume that the background cosmological constant ($\Lambda$) is given and determine the expectation value of the scalar field ($m$) or one can assume that the vacuum value of the scalar field is given and $\Lambda$ is to be determined. First consider the former case, then  one has
\begin{equation}
m^2_{\pm}= \frac{ 2 \sigma \Lambda}{\nu} \pm \frac{|\Lambda|}{\nu}\sqrt{ 4 \sigma^2 + \nu}.
\end{equation}
Since, as discussed above, the Newton's constant is fixed as $\kappa =m^{-1/2}$ , one must have $m>0$, therefore the negative root is not allowed for any sign of $\Lambda$.  The mass of the graviton \cite{BHT2,cubic} is also fixed as
\begin{equation}
M^2_g = -\sigma m_+^2 + \frac{\Lambda}{2} \hskip 2 cm  
\end{equation}
For the unitarity of the theory in dS the Higuchi bound \cite{hig} $M^2_g \ge  \Lambda >0$ must be satisfied and for unitarity in AdS Breitenlohner-Freedman bound \cite{bf} $M^2_g \ge  \Lambda $ must be satisfied \cite{BHT2,cubic}. Since these two forms are formally equal, one has
\begin{equation}
-4 \mbox{sign}(\Lambda) - 2 \sigma \sqrt{ 4+ \nu} \ge \mbox{sign}(\Lambda) \nu
 \end{equation}
where $ \mbox{sign}(\Lambda) = \Lambda/|\Lambda|$ and we have used $\sigma^2=1$. For $ \Lambda >0$,  one must have $\sigma=-1$ . For $\Lambda < 0$, both signs of $\sigma$ are allowed. 

Let us now consider the other case when the vacuum expectation value of the scalar field is assumed to be known. Then the cosmological constant is determined as 
\begin{equation}
\Lambda_{\pm}= m^2 \Big [ -2 \sigma \pm \sqrt{ 4+\nu } \Big ]. 
\end{equation}
Unitarity discussion of this case follows exactly like the one in \cite{BHT2,cubic} with the added restriction that $\nu$  is positive. Hence we do not repeat it here. 

The case when $\nu=0$ is also interesting: One has $\Lambda = -4 \sigma m^2$.   Therefore $\sigma =-1$ is allowed in dS and $ \sigma =+1$ in AdS. Suppose the expectation value of the scalar field is given, then again, Newton's constant is fixed as $m^{-1/2}$ and the graviton mass is given as $M^2_g= -3 \sigma m^2$. Higuchi bound is not satisfied therefore the theory is not unitary in dS but it is unitary in AdS. 

Now we turn to the flat vacuum of the theory which is quite subtler than the (A)dS case and our arguments will be heuristic.  It is clear that for $\Lambda =0$ in (\ref{vac}), $m$ is zero, namely, around the flat vacuum, the Weyl symmetry of the Lagrangian is  intact. The quickest way to remedy this problem is to add an explicit mass term in the Lagrangian for the scalar field and work out the details. Instead of this, let us consider the case when the symmetry is broken by radiative corrections as was shown by Coleman and Weinberg \cite{coleman} in the massless $\Phi^4$ theory. Their computation was in flat background, which we shall also work with. [Of course, one should consider the effects of gravity and the Weyl gauge field in the quantum loops, but this computation is highly non-trivial and will only change the numerical values (see the discussion below) in the computation and so it is not necessary for our main purpose of showing the {\it{ existence}} of symmetry breaking.] There is another problem: Coleman-Weinberg's computation was in four dimensions but we need the three dimensional computation for the $\nu \Phi^6$ theory. This computation was carried out in \cite{tantekin1, tantekin2}, where is was shown that, after a rather tedious renormalization and regularization procedure, at the two loop level, the effective scalar potential becomes
\begin{equation}
V_{eff}= \nu(\mu) \Phi^6 + \frac{ 7 \hbar^2}{ 120 \pi^2} \nu(\mu)^2 \Phi^6 \Big ( \ln {\frac {\Phi^4}{\mu^2} } -\frac{49}{5} \Big),
\label{eff}
 \end{equation}    
where $\mu$ is the renormalization scale and we have kept the Planck constant to show that  the symmetry is broken at the two loop level (unlike the 1 loop result in four dimensions).  
It is clear that the minimum of the potential (\ref{eff}) is away from $\Phi =0$, hence the symmetry is broken and the desired dimensionful parameter that we seek for Weyl symmetry breaking is provided by the cut-off (renormalization scale) in the quantum theory.  But there is a caveat here \cite{tantekin1,tantekin2}, the minimum is at a point where the perturbation theory breaks down, that is when $\nu(\mu)$ is large (which also happens in the four dimensional theory).  But, this will be presumably be remedied once the gauge field is taken into account
 (this is exactly what happens both in three \cite{tantekin1,tantekin2} and four dimensions \cite{coleman}.) In any case, the Weyl symmetry of the classical Lagrangian will not survive quantization, which is the relevant point  in our discussion.  

Summing up this section, we have constructed the Weyl invariant extension of NMG and have shown that NMG  appears at the Weyl-non-invariant vacuum of the extended theory both in (A)dS and flat backgrounds. Next we briefly discuss  the Weyl invariant versions of some other gravity models,

 \section{Weyl-Invariant Einstein-Gauss-Bonnet and Born-Infeld theories} 
The generic Weyl-invariant quadratic gravity is defined by the action [which can be augmented to the Weyl-invariant Einstein-Hilbert action (\ref{eh}) ] 
\begin{equation}
 \tilde{S}_{quadratic}= \int d^n x \sqrt{-g}\,\, \Phi^{\frac{2(n-4)}{n-2}} \Big [ \alpha \tilde{R}^2+\beta \tilde{R}^2_{\mu\nu}+\gamma \tilde{R}^2_{\mu\nu\rho\sigma} \Big ],
\end{equation}
where the explicit form of the curvature square terms read as
\begin{equation}
\begin{aligned}
 \tilde{R}^2= &R^2-4(n-1)R(\nabla\cdot A)-2(n-1)(n-2)R A^2+4(n-1)^2(\nabla \cdot A)^2 \\
& \quad +4(n-1)^2(n-2)A^2(\nabla \cdot A)+(n-1)^2 (n-2)^2 A^4,
\end{aligned}
\end{equation}

 \begin{equation}
\begin{aligned}
 \tilde{R}^2_{\mu\nu}= R^2_{\mu\nu}&-2(n-2)R^{\mu\nu}\nabla_\nu A_\mu -2 R(\nabla \cdot A)+2(n-2)R^{\mu\nu}A_\mu A_\nu \\
&-2(n-2)RA^2+F_{\mu\nu}^2-2(n-2)F^{\mu\nu}\nabla_\nu A_\mu \\
& +(n-2)^2 (\nabla_\nu A_\mu)^2+(3n-4)(\nabla .A )^2-2(n-2)^2 A_\mu A_\nu \nabla^\mu A^\nu \\
&+(4n-6)(n-2)A^2 (\nabla \cdot A)+(n-2)^2 (n-1)A^4,
\end{aligned}
\end{equation}

\begin{equation}
 \begin{aligned}
  \tilde{R}^2_{\mu\nu\rho\sigma} =R^2_{\mu\nu\rho\sigma} &-8 R^{\mu \nu}\nabla_\mu A_\nu+8 R^{\mu \nu}A_\mu A_\nu-4 R A^2 + n F_{\mu \nu}^2
+4(n-2)(\nabla_\mu A_\nu)^2 \\
&+4(\nabla \cdot A)^2+8(n-2)A^2 (\nabla \cdot A)-8(n-2) A_\mu A_\nu \nabla^\mu A^\nu+2(n-1)(n-2)A^4 .
 \end{aligned}
\end{equation}
From these expressions, one could study any Weyl-invariant quadratic theory.  Here, we would like to point out the specific Weyl-invariant Gauss-Bonnet combination , which gives a remarkable result in three dimensions.  In generic dimensions one has 
\begin{equation}
\begin{aligned}
 \tilde{R}^2_{\mu\nu\rho\sigma}-4\tilde{R}^2_{\mu\nu}+\tilde{R}^2&=R^2_{\mu\nu\rho\sigma}-4 R^2_{\mu\nu}+R^2+8(n-3)R^{\mu\nu}\nabla_\mu A_\nu  
-8(n-3)R^{\mu\nu} A_\mu A_\nu \\
& \quad-2(n-3)(n-4)R A^2-(3n-4)F_{\mu\nu}^2-4(n-2)(n-3)(\nabla_\mu A_\nu )^2 \\
& \quad +4 (n-2)(n-3)(\nabla \cdot A)^2 +4 (n-2)(n-3)^2 A^2 (\nabla \cdot A) \\
& \quad + 8 (n-2)(n-3)A_\mu A_\nu \nabla^\mu A^\nu-4(n-3)R (\nabla\cdot A) \\
& \quad +(n-1)(n-2)(n-3)(n-4)A^4   .           
\end{aligned}
\label{gb} 
\end{equation}
For $n=3$, Gauss-Bonnet term constructed from the metric tensor alone [the first 3 terms on the right hand side of (\ref{gb})] identically vanishes and one has 
\begin{equation}
\tilde{R}^2_{\mu\nu\rho\sigma}-4\tilde{R}^2_{\mu\nu}+\tilde{R}^2 = - 5 F_{\mu \nu}^2.
\end{equation}
Therefore, the Weyl-invariant EGB  theory reduces just to the Weyl invariant Einstein-Hilbert theory with a dynamical Weyl gauge field and a scalar. 

Finally, let us construct the Weyl-invariant version of the Born-Infeld extension of NMG \cite{BINMG}, whose action reads
\begin{equation}
 S_{BINMG}=-\frac{4 m^2}{\kappa^2} \int d^3 x \Big [ \sqrt{-\mbox{det} \Big (g+\frac{\sigma}{m^2} G \Big )}-(1-\frac{\lambda}{2})\sqrt{-g} \Big ],
\end{equation}
where the matrix $G$ is built from the Einstein tensor $ G_{\mu\nu}=R_{\mu\nu}-\frac{1}{2}g_{\mu\nu}R $. The $ S_{BINMG} $ reduces to $ S_{NMG} $ upon use of the 
small curvature expansion at $O(R^2)$
\begin{equation}
 \sqrt{(1+A )}=1+\frac{1}{2}\mbox{Tr}A+\frac{1}{8}(\mbox{T}rA)^2-\frac{1}{4}\mbox{Tr}(A^2)+O \Big (A^3\Big),
\end{equation}
 and to the theories obtained via AdS/CFT at $O(R^3)$ and beyond \cite{Sinha,cfunc,paulos}. Moreover, BINMG naturally appears as the exact, that is to all orders, counter-terms in the boundary of AdS$_4$ \cite{sinha2}. To get its Weyl-invariant version, we define the Weyl-invariant Einstein tensor 
\begin{equation}
 \tilde{G}_{ \mu \nu}= \tilde{R}_{\mu\nu}-\frac{1}{2} g_{\mu\nu} \tilde{R}.
\end{equation}
Then we have 
\begin{equation}
 S_{BINMG}=-4  \int d^3 x \Big [ \sqrt{-\mbox{det} \Big (\Phi^4 g+\sigma \tilde{G} \Big )}-(1-\frac{\lambda}{2})\sqrt{- \Phi^4 g} \Big ].
\end{equation}
Note that we have included the scalar potential here.  Expansion of the determinant in terms of the traces yields
\begin{equation}
 \begin{aligned}
\sqrt{-\mbox{det} \Big (\Phi^4 g+\sigma \tilde{G} \Big )}=\sqrt{-\mbox{det} \Big (\Phi^4 g } \Big ) \bigg ( 1- \frac{1}{2}\Phi^{-4} \tilde{R}^{\mu \nu} & \bigg [-g_{\mu \nu} + \Phi^{-4} \Big (\tilde{R}_{\mu \nu}-\frac{1}{2} g_{\mu \nu}\tilde{R} \Big ) \\ 
&+\frac{2}{3} \Phi^{-8} \Big (\tilde{R}_{\mu \rho }\tilde{R}^\rho{_\nu}-\frac{3}{4} \tilde{R}\tilde{R}_{\mu \nu}+\frac{1}{8}g_{\mu \nu} \tilde{R}^2 \Big ) \bigg ] \bigg )^{1/2} ,     
 \end{aligned}
\end{equation}
which is exact up to this point. From this expression, one can construct Weyl-invariant theories at any order in the curvature by doing a Taylor series expansion in the curvature. 

\section{Conclusions}

We have constructed a Weyl invariant extension of New Massive Gravity and have shown that the vacuum of the theory breaks Weyl symmetry and therefore, around the vacuum, the first order expansion  is just NMG with a fixed Newton's constant. Hence the mass of the graviton comes from the symmetry breaking in complete analogy with the Higgs mechanism in quantum field theory.  We have also discussed how symmetry breaking takes place in AdS and flat backgrounds: In the former, the classical field equations break the symmetry and the scalar field develops a non-zero expectation value, while in the latter, symmetry is broken at the two loop level.  We have also given the Weyl-invariant extensions of the generic quadratic models in $n$ dimensions and noted that the Weyl-invariant version of the Einstein-Gauss-Bonnet theory reduces to the Weyl-invariant Einstein-Maxwell theory with a scalar field in three dimensions.  Finally, we have given the Weyl-invariant extension of the Born-Infeld gravity in three dimensions.  Details of these models need to be worked out. It would be interesting to cast the 
Weyl-invariant theories in the elegant tractor formalism presented in \cite{waldron}. In this work, besides constructing the Weyl invariant qudratic gravity theories, we have studied the vacua of the Weyl-invariant New Massive Gravity and the unitary spin-2 excitations about these vacua in some detail. Further work is needed to understand the stability of the vacua against  the scalar and  gauge-field excitations of the theory.

\section{Acknowledgments}

B.T.  is supported by the T{U}B\.{I}TAK Grant No. 110T339, and
METU Grant BAP-07-02-2010-00-02.  We would like to thank Ibrahim Gullu and Tahsin Cagri Sisman for useful discussions.

\end{document}